\documentclass[12pt]{article}
\usepackage{graphics}
\usepackage{latexsym}

\newcommand{\Tr}{{\rm Tr}}

\newcommand{\bare}{{\rm bare}}
\newcommand{\reno}{{\rm reno}}
\newcommand{\daggerC}{{\dagger}}

\begin{document}

\begin{flushright}
 \begin{minipage}[b]{43mm}
  DPNU-99-03\\
  hep-th/9904193\\
  Revised on August 2000
 \end{minipage}
\end{flushright}

\renewcommand{\thefootnote}{\fnsymbol{footnote}}
\begin{center}
 {\LARGE\bf Effective Theoretical Approach to Backreaction of
 	    the Dynamical Casimir Effect in 1+1 Dimensions}\\
 \vspace{5mm}
 {\large Yukinori Nagatani}\footnote
 {E-mail: nagatani@yukawa.kyoto-u.ac.jp}\\[2mm]
 {\it Yukawa Institute for Theoretical Physics, Kyoto University,\\
      Sakyo-ku, Kyoto 606-8502, Japan}\\[4mm]
 and\\[4mm]
 {\large Kei Shigetomi}\footnote
 {E-mail: shige@eken.phys.nagoya-u.ac.jp}\\[2mm]
 {\it Department of Physics, Nagoya University,\\
      Nagoya 464-8602, Japan}
\end{center}
\vspace{0mm}

\begin{abstract}
 We present an approach to studying the Casimir effects
 by means of the effective theory.
 An essential point of our approach is replacing the mirror separation
 into the size of space $S^1$ in the adiabatic approximation.
 It is natural to identify the size of space $S^1$ with the scale factor
 of the Robertson-Walker-type metric.
 This replacement simplifies the construction of a class of effective models
 to study the Casimir effects.
 To check the validity of this replacement
 we construct a model for a scalar field coupling to the two-dimensional
 gravity and calculate the Casimir effects
 by the effective action for the variable scale factor.
 Our effective action consists of the classical kinetic term of the
 mirror separation and the quantum correction derived by the 
 path-integral method.
 The quantum correction naturally contains both the Casimir energy term and
 the back-reaction term of the dynamical Casimir effect,
 the latter of which is expressed by the conformal anomaly.
 The resultant effective action describes
 the dynamical vacuum pressure, i.e., the dynamical Casimir force.
 We confirm that the force depends on the relative velocity of the mirrors,
 and that it is always attractive and stronger than the static Casimir
 force within the adiabatic approximation.\\[5mm]
 PACS number(s): 12.20.Ds, 03.65.Ca, 03.70.+k, 11.15.Kc
\end{abstract}

\newpage
\section{INTRODUCTION}\label{intro.sec}


The Casimir effect originally suggested in 1948
has been generally regarded as the contribution of a nontrivial geometry
on the vacuum fluctuations of quantum electromagnetic
fields~\cite{Casimir,Boyer}.
The change in the vacuum fluctuations caused by the change of geometry
appears as a shift of the vacuum energy
and a resulting vacuum pressure.
For a standard example,
when we insert two perfectly conducting parallel plates
into the free space $R^3$,
the plates are attracted towards each other~\cite{Casimir},
although being uncharged.
This attractive force
is experimentally confirmed by Sparnaay in 1958~\cite{Sparnaay}
and recently more precise measurements have been provided~\cite{Lamoreaux}.


The dynamical Casimir effect suggests
that the nonuniform accelerative relative motion of the boundaries
(perfectly conducting plates or mirrors)
excites the electromagnetic field and
promotes virtual photons from the vacuum into real photons
~\cite{DCE_Moore,DCE_FD,DCE_Sassaroli,DCE_Dodonov,DCE_some}.
The works on the dynamical Casimir effect are
pioneered by Moore~\cite{DCE_Moore}
and have progressed by many
authors~\cite{DCE_FD,DCE_Sassaroli,DCE_Dodonov,DCE_some}.
Moore studied the quantum theory of a massless scalar field 
in the one-dimensional cavity bounded by moving mirrors,
and evaluated the number of photons 
created by the exciting effect of the moving mirrors.
In his approach, 
the boundary condition on the scalar field is replaced with
the simple equation, referred to as the Moore's equation,
which describes the constraint on the conformal transformation
of the coordinate.
His approach has been popularly used to investigate the 
problems relating to the $(1+1)$-dimensional dynamical Casimir effect.
For a well-known example Fulling and Davies calculated
the energy-momentum tensor with the Moore's equation,
and showed the existence of the radiation from the moving
mirrors~\cite{DCE_FD}.


The dynamical Casimir effect occurs even in the adiabatic approximation.
Indeed, we can hardly handle
the configurations except for the adiabatic deformations.
Here, {\it adiabatic} means the absence of mixings
among the different energy levels of the system
during the modulation of the mirror separation.
In other words the relative velocity of the mirrors is much
smaller than the velocity of light.
Especially Sassaroli {\it et al.} succeeded
in evaluating the number of photons
produced by the adiabatic motion of the mirrors
in $1+3$ dimensions~\cite{DCE_Sassaroli}.
They used the Bogolubov transformation among the creation
and annihilation operators of photon
in order to describe the particle production.


The similar phenomena of the particle production also have been predicted
in a variety of general-relativistic
situations~\cite{GRPCreation,HawkingRadiation,Davis}.
Such phenomena include the Hawking radiation
from black holes~\cite{HawkingRadiation},
the domain-wall activity in cosmology,
and the high-speed collision of atomic nuclei~\cite{Davis}.
Although these phenomena are interesting,
the dynamical Casimir effect has not yet been experimentally confirmed.


If moving mirrors create radiation,
the mirrors experience a radiation-reaction force.
Several authors have discussed this subject
within the adiabatic approximation.
Dodonov {\it et al.} showed the existence of
the additional negative frictional force 
besides the static Casimir force in the one-dimensional cavity
by using Moore's equation~\cite{DCE_Dodonov}.

%
The advantage of Moore's approach is the properties:
the theory does not need to possess the Hamiltonian or the Lagrangian
to describe the time evolution of the field.
However, it seems difficult to apply Moore's approach to
study the Casimir effects and its backreaction
in $1+2$ or $1+3$ dimensions
because the boundary condition of the one-dimensional space
plays a crucial role in his approach.


In this paper
we present an effective-theoretical approach to studying the Casimir
effects in $1+1$ dimensions.
Our approach, making use of the action,
is considered to be applicable to study the 
Casimir effects and its back reaction also in the higher dimensions.
In general
the existence of the moving boundaries (mirrors) makes it difficult to
construct the Hamiltonian or the Lagrangian describing the system,
since the relative motion of the boundaries (mirrors) mixes
one energy level of the system with the others.
However, we note that
the adiabatic motion allows us to neglect this boundary effect:
we do not need the boundaries.
So we replace the spatial configuration $D^1$ into $S^1$
in the adiabatic approximation.
The motion of the cavity size is described by varying the radius
of $S^1$ in time.
Furthermore, we can naturally identify the size of space $S^1$
with the scale factor of the Robertson-Walker-type metric.
That is, the mirror separation is described by the scale factor.
The time evolution of the scale factor
can be regarded as the space-time $R\times S^1$ with gravity.
For the sake of the replacement from $D^1$ into $S^1$,
we can study the Casimir effects from the viewpoint
of the effective theory.
The construction of the model with the replacement
is very simple and general,
so that it is easy to apply our approach to
more realistic models in the higher dimensions
by replacing the space 
$D^1 \times R^n$ into $S^1\times R^n$.


To check the validity of our replacement,
we construct a scalar model and calculate the Casimir effects.
%
%
As is usual our model makes use of the conformal symmetry property of
the two-dimensional theory of massless fields.
In our model of the cavity-system
the classical action is constructed by
the classical kinetic term of the
mirror separation and the Polyakov action.
The Polyakov action describes the massless scalar field
minimally coupling to the two-dimensional gravity.
The classical action is simple and general,
so the structure of the model, e.g., symmetry, is easily visible.
We carry out the path integral on the scalar field,
and obtain the effective action for the mirror separation.
The calculation of the path integral is rather complicated;
however, it can be exactly performed.
The effective action consists of
the classical kinetic term of the mirror separation
and the quantum correction terms.
The quantum correction takes a well-known form,
which consists of the static Casimir energy term and
the conformal anomaly term.
The conformal anomaly term represents the back reaction of the dynamical
Casimir effect.
The effective action finally leads to the dynamical vacuum pressure
depending on the relative velocity of the mirrors.

Our approach also gives an explanation for the origins of the Casimir
effects in terms of the effective theory:
the Casimir effects are caused by the change of field configuration in
the vacuum instead of the existence of the boundaries.


The paper is organized as follows.
In Sec. \ref{model.sec}
we provide the general description of our model
and the definition of the effective action.
In spite of the simplicity of our model,
the calculation of the effective action is rather complicated.
We show the calculation in detail in the following two sections.
In Sec. \ref{partition.sec}
the Casimir energy is shown to be derived from
the partition function part in the effective action.
In Sec. \ref{conf.sec}
the conformal anomaly part in the effective action is calculated,
and obtained the back-reaction term of the dynamical Casimir effect.
In Sec. \ref{back.sec}
the back reaction of both the Casimir effects in our model is
investigated, and the dynamical vacuum pressure is derived.
Section \ref{summary.sec} is devoted to conclusions and discussions.
In Appendix A
the conformal anomaly is induced
by means of the Fujikawa method~\cite{Fujikawa}
and in Appendix B
another path-integral calculation on the Casimir energy are shown.

\section{SCALAR MODEL FOR CASIMIR EFFECTS}\label{model.sec}

\begin{figure}[htbp]%
\begin{center}%
 \includegraphics{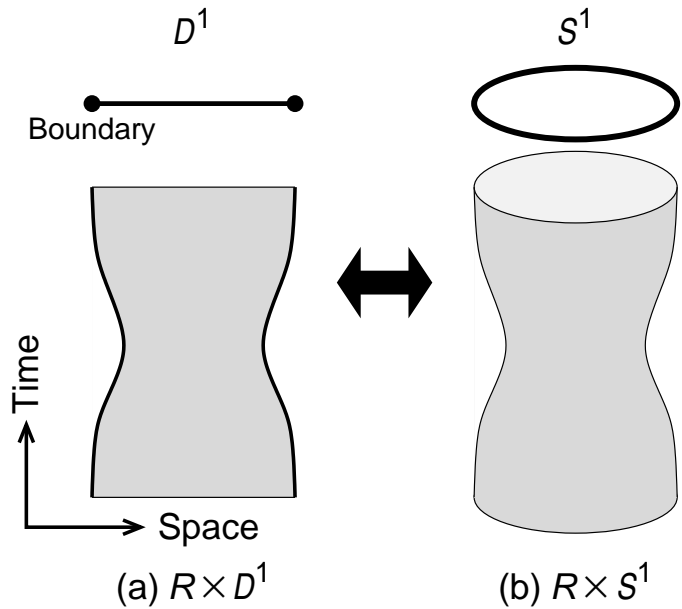}
 \caption{Space times for the $(1+1)$-dimensional Casimir effects.
 (a)
 One-dimensional space with two boundaries
 (one-dimensional disk $D^1$)
 as the cavity between two moving ``mirrors.''
 The scalar field satisfies the fixed boundary condition
 on the edges.
 (b)
 The space $S^1$ which is adiabatically equivalent for the
 scalar field  to the geometrical configuration (a).
 The periodic boundary condition is imposed on the field. }%
 \label{SpaceTime.eps}%
\end{center}%
\end{figure}

%
The steps for constructing our model are as follows:
%
%
For the purpose of describing the Casimir effects
in the one-dimensional cavity
and the reaction received by the moving mirrors,
we consider a massless scalar field
in the one-dimensional finite space with two boundaries,
i.e., one-dimensional disk $D^1$ [see Fig. \ref{SpaceTime.eps}(a)].
That is, we consider the scalar field between two moving
``mirrors.''
The size of $D^1$ is a dynamical variable,
and we assume that the size receives
all the back reaction of the Casimir effects.


The motion of the boundaries generally mixes
the energy levels of the system.
However, when the motion of the mirror separation is adiabatic,
there are no transitions among the energy levels~\cite{DCE_Sassaroli}.
Because of this absence of the transitions
we can neglect the existence of the boundaries.
This implies that each adiabatic Hamiltonian in the space $D^1$
is the same as that in the space $S^1$ except for the overall factor.
We replace the spatial configuration $D^1$ with $S^1$
in the adiabatic approximation [see Fig. \ref{SpaceTime.eps}(b)].
In the space $S^1$ the scalar field is required
to satisfy the periodic boundary condition
rather than the fixed boundary condition.
Accordingly the energy levels of the adiabatic oscillation modes
in the replaced system are two times as those in the original system.
We can naturally regard the size of $S^1$ as the scale factor of the
Robertson-Walker-type metric.
We define the Robertson-Walker-type metric
on the space-time $R\times S^1$:
\begin{eqnarray}
ds^2 &=& -dt^2 + D^2(t)dx^2 \qquad (0\le x \le a),
\end{eqnarray}
where a dimensional constant $a$ is the standard space size and 
the scale factor $D(t)$ is the dimensionless magnification rate.
It should be noticed that the mirror separation is replaced with
the scale factor of the metric.

With the help of this replacement, the model
in the two-dimensional gravity is applicable to our model.
The mirror separation has finite reduced mass $m$ and classically obeys
free motion.
Then the classical action of our model to describe the system
consists of both the classical kinetic term for
the scale factor and the Polyakov action,
\begin{eqnarray}
 S[D,\phi] &\equiv& \frac{1}{2\pi} \int dt
  \frac{m}{2} a^2 \dot{D}^2(t)
  \;+\; \frac12S_{\rm Polyakov}[g_{\mu\nu}(D),\,\phi],
\end{eqnarray}
where
\begin{eqnarray}
S_{\rm Polyakov}[g_{\mu\nu},\,\phi] &=& -\frac1{2\pi}\int d^2x\:
\sqrt{-g}\:\frac12\:g^{\mu\nu}
\partial_{\mu}\phi\partial_{\nu}\phi.
\label{Polyakov.eq}
\end{eqnarray}
The Polyakov action is invariant under both the general coordinate
transformation and the Weyl transformation.
This property is referred to as the conformal symmetry.
We can always rewrite the metric into the conformal flat form
by the general coordinate transformation:
\begin{eqnarray}
 ds^2
  &=& -dt^2 + D^2(t)dx^2 
  \;=\; - C(\eta)\left(d\eta^2 - dx^2\right)
  \;=\; g_{\mu\nu} \; dx^\mu dx^\nu,
  \label{RWMetric.eq}
\end{eqnarray}
where we have introduced a new coordinate $\eta$ such that
$d\eta\equiv dt/D(t)$ and $C(\eta)\equiv D^2[t(\eta)]$.
After performing the Weyl transformation
$g_{\mu\nu}\to C^{-1}(\eta) g_{\mu\nu}$,
we have the $D(t)$-independent flat metric,
\begin{eqnarray}
 ds^2
  &=& -d\eta^2 + dx^2
  \;=\; \eta_{\mu\nu} \; dx^\mu dx^\nu.
  \label{FlatMetric.eq}
\end{eqnarray}
This implies that any deformation of the space size does not affect
the classical action.
But once we quantize the scalar field,
the conformal anomaly appears in general.
The quantum effects lead to the motion of the scale factor,
i.e., the motion of the mirror separation.


We use a path-integral formulation to evaluate the motion of $D(t)$
as the back reaction of the Casimir effects.
We use the background field method, in which the metric is treated as
a classical field and the scalar field is quantized.
We obtain the effective action for $D(t)$
by integrating out the scalar field.
The effective action for the metric, $S_{\rm eff}[D]$, is given by
\begin{eqnarray}
 e^{iS_{\rm eff}[D]}
  &\equiv&
  \int{\cal D}\phi \; e^{iS[D,\,\phi]}\\
 &=& 
  e^{i \frac{1}{2\pi} \int dt
  \frac{m}{2} a^2 \dot{D}^2(t) \;+\; i\frac12\Gamma[g_{\mu\nu}(D)]},\\
 e^{i\Gamma[g_{\mu\nu}]}
  &\equiv&
  \int{\cal D}\phi \; e^{iS_{\rm Polyakov}[g_{\mu\nu},\,\phi]}.
  \label{EffAction1.eq}
\end{eqnarray}
In order to calculate the effective action
for the evolving metric (\ref{RWMetric.eq}),
we perform the conformal transformation on
the effective action (\ref{EffAction1.eq})
from the evolving metric (\ref{RWMetric.eq})
to the flat metric (\ref{FlatMetric.eq}):
$g_{\mu\nu}\to e^{2\alpha} g_{\mu\nu}
 = C^{-1}(\eta) g_{\mu\nu} = \eta_{\mu\nu}$.
By means of the Fujikawa method~\cite{Fujikawa}
this conformal transformation picks up the conformal anomaly
as a Jacobian factor from the path-integral measure
in the effective action (\ref{EffAction1.eq}):
\begin{eqnarray}
 e^{i\Gamma[g_{\mu\nu}]}
 &=&
 \exp\left[-\frac{i}{2}\int d^2x\,\alpha(x)
      \sum_n\varphi^{\daggerC}_n(x)\,\varphi_n(x)
     \right]
  e^{i\Gamma[\eta_{\mu\nu}]} \quad ,
  \label{EffAction2.eq}
\end{eqnarray}
where the parameter of the conformal transformation $\alpha(x)$
is chosen as $\alpha(x) = - \frac12 \ln C(\eta)$.
$\{\varphi_n(x)\}$ is a complete set which consists of
the eigenfunctions of the Hamiltonian (see Appendix A).
The first exponential factor in Eq. (\ref{EffAction2.eq}) is the
conformal anomaly, and the second factor is the partition function for
the free scalar field in the space $S^1$.

\section{CASIMIR ENERGY IN SPACE $S^1$}\label{partition.sec}

We will see that $\Gamma[\eta_{\mu\nu}]$ induces the Casimir energy
as the vacuum energy by evaluating the partition function for the free
scalar field.
Let us calculate the Euclidean partition function
\begin{eqnarray}
 Z_E
  &\equiv&
   e^{-\Gamma_E[\eta_{\mu\nu}]}
  \;=\;
   \int{\cal D}{\phi}\,
   \exp \left[
	 -\frac1{2\pi} \int_{-\infty}^{\infty} dx^2 \int_0^a dx^1
	 \frac12 \, \partial\phi \, \partial\phi
	\right],
\label{partition.eq}
\end{eqnarray}
where
we have defined the imaginary time variable $x^2 \equiv i\eta$,
and
have used the Euclidean inner product
$\partial\phi \partial\phi
 \equiv \delta^{\mu\nu}\partial_\mu\phi \partial_\nu\phi$.
Since the free Lagrangian is quadratic in terms of $\phi$,
this integration can be performed formally, and obtains 
\begin{eqnarray}
 \ln Z_E
  &=&
	-\frac12 \Tr \ln (\partial^2) 
 \;=\;
	-\frac12 \int d^2x \; \langle x | \ln\partial^2 | x \rangle.
\end{eqnarray}
In the momentum representation
the spatial component of the momentum
is discretized in the form $(2\pi n / a)$
for arbitrary integers $n$ due to the compactness of the space,
\begin{eqnarray}
 \ln Z_E
 &=&
 	-\frac12 \, \frac1{(2\pi)^2}
	\int d^2x \int \frac{dk}{2\pi} \, \frac1a \,
	\sum_n \; \ln\left[ k^2 + \left(2\pi n / a\right)^2 \, \right]
 \;\equiv\;
	-\frac1{(2\pi)^2}\int d^2x f_{\bare},
	\label{Fbare.eq}
\end{eqnarray}
where $f_{\bare}$ is a bare Euclidean free-energy density for the
massless field.
Since the integration over $k$ makes $f_{\bare}$ divergent,
we introduce mass $M$ of the scalar field
to regularize $f_{\bare}$~\cite{Ganguly},
then the integrand is changed as
\begin{eqnarray}
 \ln\left[k^2 \:+\: \left(2\pi n / a\right)^2\,\right]
  &\to&
 \ln\left[k^2 \:+\: \left(2\pi n / a\right)^2 \:+\: M^2\right].
\end{eqnarray}
Employing the indefinite integral of $M$,
we can write
\begin{eqnarray}
 f_{\bare} 
  &=& \frac 12\int \frac{dk}{2\pi}\frac 1a
  \int dM^2 \sum_n \frac 1{k^2 + \left(2\pi n / a\right)^2 + M^2}.
\end{eqnarray}
The sum over $n$ can be performed in the expression
\begin{eqnarray}
 f_{\bare}
  &=&
	\frac 12 \int \frac{dk}{2\pi}
	\int d\omega_k
	\left(1 + 2\sum_{n=1}^{\infty} e^{-na\omega_k} \right),
	\label{Fbare2.eq}
\end{eqnarray}
where we have employed $k$ and $\omega_k\equiv\sqrt{k^2 + M^2}$
as independent parameters instead of using $k$ and $M$,
and have used the identity
\begin{eqnarray}
 \sum_{n=-\infty}^{\infty}\frac 1{n^2+b^2}
 &=& \frac{\pi}b \coth\pi b
 \;=\; \frac{\pi}b \: \left[1 + \frac 2{e^{2\pi b} -1}\right]
 \;=\; \frac{\pi}b \:
 \left[1 + 2\sum_{n=1}^{\infty} e^{-2\pi nb}\right].
\end{eqnarray}
Since the first term of Eq. (\ref{Fbare2.eq})
indicates the contribution of infinite volume of space time and
clearly diverges, we renormalize it as a cosmological term.
The second term is relevant for the free-energy density,
namely, renormalized free-energy density,
\begin{eqnarray}
 f_{\reno}
  &\equiv&
	-\frac1a \int \frac{dk}{2\pi} \:
	\sum_{n=1}^{\infty} \: \frac 1n \: e^{-na\omega_k}.
\end{eqnarray}
With the identity
\begin{eqnarray}
 e^{-na\omega_k}
  &=&
	\frac{1}{\sqrt{\pi}} \int_0^{\infty} dt\;t^{-1/2}
	\exp \left[ -t - \frac{\left(na\omega_k\right)^2}{4t} \right],
\end{eqnarray}
we perform the integration over $k$, and obtain 
\begin{eqnarray}
 f_{\reno}
  &=&
	-\frac{M}{\pi a} \: \sum_{n=1}^{\infty} \: \frac1n K_{-1} (naM).
  \label{Freno1.eq}
\end{eqnarray}
Here $K_\nu(z)$ is the modified Bessel function
\begin{eqnarray}
 K_{\nu}(z)
  &=&
  \frac 12\left(\frac z2\right)^{\nu}
  \int_0^{\infty} \frac{dt}{t^{\nu+1}} \exp \left[-t-\frac{z^2}{4t}\right].
  \label{Bess.eq}
\end{eqnarray}
The free-energy density for the massless field is obtained
by taking the limit $M \to 0$.
In this limit
we can use the property of the Bessel function,
$K_{-1}(z) \approx 1/z$ for small $z$,
and the free-energy density (\ref{Freno1.eq}) becomes
\begin{eqnarray}
f_{\reno}
 &=& -\frac 1{\pi a^2}\sum_{n=1}^{\infty}\frac 1{n^2}
 \;=\; -\frac{\pi}{6a^2}.
 \label{Freno2.eq} 
\end{eqnarray}
The Euclidean partition function is derived by substituting
Eq. (\ref{Freno2.eq}) into Eq. (\ref{Fbare.eq}).
After performing the spatial integration,
and going back to the Minkowski space with $x^2 = i\eta$,
we obtain
\begin{eqnarray}
 \Gamma[\eta_{\mu\nu}]
  &=&
  	\frac{1}{i} \ln Z
  \;=\;
  	\frac 1{2\pi} \int_{-\infty}^{\infty}dt \;
	\frac 1{12} \frac 1{aD(t)},
	\label{PartitionFuncResult.eq}
\end{eqnarray}
where we have used the relation $d\eta = dt/D(t)$.
It should be noticed that $-1/(12aD)$
is the Casimir energy in $1+1$ dimensions,
and is caused not by the existence of the boundary
but by the compactness of the space.

\section{CONFORMAL ANOMALY IN SPACE-TIME $R\times S^1$}\label{conf.sec}

In this section 
the effective action for the metric $\Gamma[g_{\mu\nu}]$
is derived by evaluating the conformal anomaly in the space-time
$R\times S^1$.
The conformal anomaly is formally expressed by the first exponent in the
right-hand side of Eq. (\ref{EffAction2.eq}).
This anomaly part appears
when the space-size $S^1$ is varying with time.
Then the anomaly part is considered to describe the back-reactional terms 
of the dynamical Casimir effect.
In the Euclidean space time with the metric
$ds^2 = \rho(x^2) \: \left[\left(dx^1\right)^2 + \left(dx^2\right)^2\right]$
the Jacobian induced from the conformal transformation
$g_{\mu\nu} \to e^{2\alpha} g_{\mu\nu}$ is
\begin{eqnarray}
 J_E
  &\equiv&
	\exp \left[-\frac 12\int_0^adx^1
	      \int_{-\infty}^{\infty} dx^2 \,
             \alpha(x^2) \sum_{n,k} \varphi^{\dagger}_{n,k}(x)\,
	     \varphi_{n,k}(x)
	     \right],
	 \label{JacobianE1.eq}
\end{eqnarray}
where $\{\varphi_{n,k}(x)\}$ is a complete set
of the eigenfunctions of the Hamiltonian operator,
\begin{eqnarray}
 \hat{H}
  &=&
	-\frac12\,\frac1{\sqrt{\rho}}\,\partial\partial\,\frac1{\sqrt{\rho}},
  \qquad
	\hat{H} \varphi_{n,k}(x)
  \;=\;
	\lambda_{n,k}^2\varphi_{n,k}(x).
	\label{Hamiltonial.eq}
\end{eqnarray}
This Jacobian will be evaluated
by using the eigenfunctions $\varphi_{n,k}(x)$
which satisfy the periodic boundary condition in the space $S^1$.

The factor $j(x) \equiv \sum_{n,k}
\varphi^{\dagger}_{n,k}(x)\varphi_{n,k}(x)$
in the Jacobian (\ref{JacobianE1.eq}) has a divergence
due to the infinite degrees of freedom of the space-time points.
In order to regularize this divergence
we introduce a cutoff parameter $M$ and 
insert the cutoff function $\exp(-\lambda_{n,k}^2 / M^2)$ into $j(x)$:
\begin{eqnarray}
 &&
	j(x) \;\equiv\;
	\sum_{n,k} \varphi^{\daggerC}_{n,k}(x) \varphi_{n,k}(x)\nonumber\\
 &\;\to\;&
	j(x) \;\equiv\; 
	\lim_{M\to\infty}
	\sum_{n,k} \varphi^{\daggerC}_{n,k}(x)
	e^{-\lambda_{n,k}^2 / M^2} \varphi_{n,k}(x)
 = 
	\lim_{{M\to\infty}}
	\sum_{n,k}
	\varphi^{\daggerC}_{n,k}(x)
	e^{-\hat{H} / M^2}
	\varphi_{n,k}(x).
	\nonumber
\end{eqnarray}
When we take 
$\varphi_{n,k}(x)=\frac1{\sqrt{a}}e^{ikx^2}e^{i(2\pi n / a) x^1}$
as the eigenfunction, 
we obtain
\begin{eqnarray}
j(x) \;=\;
	\lim_{M\to\infty} \frac1a
	\sum_{n=-\infty}^{\infty}
	\int_{-\infty}^{\infty} \frac{dk}{2\pi}
	\exp \left[
	      \frac 1{2 M^2}
	      \left(
	       -\frac{k^2+\left(2\pi n / a\right)^2}{\rho}
	       + 2\frac {ik}{\sqrt\rho}\partial_2\frac 1{\sqrt\rho}
	       + \frac 1{\sqrt\rho}\partial_2^2\frac 1{\sqrt\rho}
	      \right)
	     \right].
	\label{jFunc1.eq}
\end{eqnarray}
Here we should note that $j(x)$ is independent of $x^1$.
Redefining $k \to Mk$,
we can write Eq. (\ref{jFunc1.eq})
with a dimensionless parameter $k$ as
\begin{eqnarray}
 j(x^2)
  &=&
   \lim_{M\to\infty} \frac{M}{a}
   \sum_{n=-\infty}^{\infty}
   \int_{-\infty}^{\infty}\frac{dk}{2\pi}
   \exp\left[
	- \frac{{k}^2+\left(2\pi n / Ma\right)^2}{2 \rho }
	+ \frac{i{k}}{M} \frac1{\sqrt\rho} \partial_2 \frac1{\sqrt\rho}
	+ \frac1{2M^2} \frac1{\sqrt\rho} \partial_2^2 \frac1{\sqrt\rho}
	\right].\nonumber\\
 &&\label{jFunc2.eq}
\end{eqnarray}
The second and the third terms in the exponent in Eq. (\ref{jFunc2.eq})
are understood as operators, e.g.,
\begin{eqnarray}
      \frac1{\sqrt\rho} \partial \frac1{\sqrt\rho}
 &=& -\frac{\partial\rho}{2\rho^2} + \frac1{\rho}\partial .
\end{eqnarray}
After expanding the integrand in terms of $M^{-1}$,
the order $M^2$ terms in Eq. (\ref{jFunc2.eq}) under integrating over $k$
and summation over $n$,
denoted as ${\cal O}(M^{2})$, diverge with the limit on $M$.
Notice that $\sum_{n=-\infty}^{\infty}\exp\left[-\frac{\left(2\pi
n/Ma\right)^2}{2\rho }\right]$ gives the contribution of ${\cal O}(M)$.
The part of ${\cal O}(M^{2})$, however, is renormalizable by adding 
a bare cosmological term to the starting Lagrangian~\cite{Fujikawa,PL}.
In this expansion
the terms in Eq. (\ref{jFunc2.eq}) including only one operator
$\frac{i{k}}{M}\frac1{\sqrt\rho} \partial_2 \frac1{\sqrt\rho}$
become ${\cal O}(M)$ because of the
existence of the dumping factor, $\exp(-{k}^2/2\rho)$.
The part of ${\cal O}(M)$ in Eq. (\ref{jFunc2.eq}) becomes zero for
symmetric integration on the odd function.
Then the next reading terms of ${\cal O}(M^{0})$ in Eq. (\ref{jFunc2.eq})
remain under the limit on $M$.
The terms of ${\cal O}(M^{0})$ in Eq. (\ref{jFunc2.eq})
consist of two kinds of contributions.
One comes from the operator
$\frac1{2M^2} \frac1{\sqrt\rho} \partial_2^2 \frac1{\sqrt\rho}$
in Eq. (\ref{jFunc2.eq}), becoming
\begin{eqnarray*}
 &&
   \frac1{Ma}
   \sum_{n=-\infty}^{\infty}
   \int_{-\infty}^{\infty}\frac{dk}{2\pi}
   \;
   \sum_{m=1}^{\infty}
   \frac1{m!}
   \left[
    	- \frac{k^2 + \left(2\pi n / Ma \right)^2}{2 \rho}
    \right]^{m-1}
   \\
 && \qquad\qquad \times
  \left\{
   	\left(
	 	\frac{m^3}{6} + \frac{m^2}{4} - \frac{m}{24}
	\right)
	\rho^{-3} (\partial_2\rho)^2
	- \frac{m^2}4 \rho^{-2} \partial_2^2 \rho
  \right\},
\end{eqnarray*}
and another comes from the two operators of
$\frac{i{k}}{M} \frac1{\sqrt\rho} \partial_2 \frac1{\sqrt\rho}$,
being
\begin{eqnarray*}
 &&
   \frac1{Ma}
   \sum_{n=-\infty}^{\infty}
   \int_{-\infty}^{\infty}\frac{dk}{2\pi}
   \;
   \sum_{m=1}^{\infty}
   \frac{(ik)^2}{m!}
   \left[
 	  - \frac{k^2 + \left(2\pi n / Ma \right)^2}{2 \rho}
   \right]^{m-2}
   \\
 && \qquad\qquad \times
  \left\{
   	\left(
	 	\frac{m^4}{8} - \frac{m^2}{4} + \frac{m}{8}
	 \right)
	 \rho^{-4}(\partial_2\rho)^2
       - \left(
	  	\frac{m^3}{6}-\frac{m^2}{4}+\frac{m}{12}
	 \right)
	 \rho^{-3}\partial_2^2\rho
  \right\}.
\end{eqnarray*}

After performing the integration over $k$,
$j(x^2)$ becomes
\begin{eqnarray}
 j(x^2)
  &=&
   \lim_{M\to\infty}
   \left[
         F(\rho) \left(\partial_2\rho\right)^2
       + G(\rho) \partial_2^2\rho
   \right],
\end{eqnarray}
where $F(\rho)$ and $G(\rho)$ are given by
\begin{eqnarray*}
 F(\rho)
  &\equiv&
  	\frac{\rho^{-5/2}}{\sqrt{2\pi}Ma}
	\sum_{n=-\infty}^{\infty}
        \exp\left[
	      - \frac12
	        \left( \frac{2\pi n}{Ma\sqrt{\rho}} \right)^2
	     \right]
	\left\{
	         \frac5{32} 
	       - \frac5{48}
	         \left( \frac{2\pi n}{Ma\sqrt{\rho}} \right)^2
	       + \frac1{96}
	         \left( \frac{2\pi n}{Ma\sqrt{\rho}} \right)^4
 	\right\},\\
 G(\rho)
  &\equiv&
  	\frac{\rho^{-3/2}}{\sqrt{2\pi}Ma}
	\sum_{n=-\infty}^{\infty}
	\exp\left[
	     - \frac12
	       \left( \frac{2\pi n}{Ma\sqrt{\rho}} \right)^2
	    \right]
	\left\{
		- \frac18    
	        + \frac1{24}
	          \left( \frac{2\pi n}{Ma\sqrt{\rho}} \right)^2
	\right\}.
\end{eqnarray*}
Under the limit on $M$ we obtain
\begin{eqnarray}
 \lim_{M\to\infty} F(\rho) &=&     \frac{\rho^{-2}}{2\pi} \frac{1}{12},
 \qquad
 \lim_{M\to\infty} G(\rho) \;=\; - \frac{\rho^{-1}}{2\pi} \frac{1}{12},
\end{eqnarray}
with the help of the definition of the Jacobi $\theta$ function
and its property:
\begin{eqnarray*}
 \theta(u,\tau)
  &\equiv&
  \sum_{l=-\infty}^{+\infty}
  \exp\left( 2\pi iul + i\pi\tau l^2 \right), \\
 &\theta(0,i\tau)&
 \;=\;
 \frac1{\sqrt\tau} \: \theta\left(0,\,\frac{i}{\tau}\right).
\end{eqnarray*}

In order to evaluate the effective action (\ref{EffAction2.eq})
with the Euclidean metric
$ds^2 = \rho(x^2) \:\times\:
 \left[\left(dx^1\right)^2 + \left(dx^2\right)^2\right]$,
we have to choose the parameter of the conformal transformation as
$\alpha(x^2) = -\frac12\ln \rho(x^2)$.
Then the Jacobian factor (\ref{JacobianE1.eq}) becomes
\begin{eqnarray}
 \ln J_E[\rho]
  &=&
  	\frac{1}{96 \pi}
	\int_0^a dx^1 \,
	\int_{-\infty}^\infty dx^2 \,
	\rho^{-2} (\partial_2 \rho)^2.
\end{eqnarray}

Now we continue back to the Minkowski Jacobian
with time evolving metric (\ref{RWMetric.eq}):
\begin{eqnarray}
 \frac{1}{i} \ln J[D]
  &=&
  	- \frac{1}{2\pi} \int dt \; \frac{a}{12} \frac{{\dot D}^2}D,
	\label{JacobianM2.eq}
\end{eqnarray}
where we have used the relations
between the Euclidean parameters and the Minkowski ones:
$x^2=i\eta$, $\rho(x^2) = C(\eta)$,
and we note that $d\eta=dt/D(t), C(\eta)=D(t)^2, \int dx=a$.

On the other hand, the well-known Polyakov-Liouville action~\cite{PL},
which is the conformal anomaly in the space-time $R^2$,
brings the same result as Eq. (\ref{JacobianM2.eq}), shown as follows.
The Polyakov-Liouville action is given by the general form:
\begin{eqnarray}
 S_{\rm PL}
   &=& - \frac{1}{96\pi} \int d^2x \: \sqrt{-g} \:
         \int d^2x' \: \sqrt{-g'}
	    R(x)\Box^{-1}(x,x') \: R(x'),
\end{eqnarray}
where $R(x)$ is the Ricci curvature.
With the form of the metric, $ds^2 = -C(\eta)\:(d\eta^2 -dx^2)$,
\begin{eqnarray}
 S_{\rm PL}
   &=& - \frac{1}{96\pi} \int_{0}^{a}dx\int d\eta \: C \:
         \ln C \: \Box  \ln C,
         \label{PL.eq}
\end{eqnarray}
and the Ricci curvature is $R(x)=-\Box\ln C$.
With the relations, $d\eta=dt/D(t)$ and $C(\eta)=D(t)^2$,
we come back to the Robertson-Walker-type metric $ds^2 = -dt^2+D(t)^2dx^2$,
and obtain the Ricci curvature in terms of $D(t)$:
\begin{eqnarray}
 R(x) &=& -\Box\ln C \;=\; \frac{2\ddot{D}}{D}.
 \label{Ricci}
\end{eqnarray}
Here we use $\Box = g^{\mu\nu}\partial_{\mu}\partial_{\nu}
= -\frac{1}{C}\partial_{\eta}^2$,
and the relation $\partial_{\eta}=D\partial_{t}$.
By substituting Eq. (\ref{Ricci}) into Eq. (\ref{PL.eq}),
$S_{\rm PL}$ is modified as
\begin{eqnarray}
  S_{\rm PL}
   &=& \frac{1}{24\pi} \int_{0}^{a}dx \int dt \: \ddot{D}\ln D.
    \label{CA.eq}
\end{eqnarray}
This result is consistent with the well-known fact that the regulated
trace of the stress tensor is proportional to the curvature.
After the partial integration, Eq. (\ref{CA.eq}) is
found to be the same as our result (\ref{JacobianM2.eq}),
which is the case of $R\times S^1$.

Finally, combining the partition function (\ref{PartitionFuncResult.eq})
and the Jacobian factor (\ref{JacobianM2.eq})
gives the effective action for the space size $D(t)$ as
\begin{eqnarray}
 \Gamma[D]
  &\equiv&
	\Gamma[g_{\mu\nu}]
  \;=\;
	\frac{1}{i} \ln J + \frac{1}{i} \ln Z
  \;=\;
	\frac 1{2\pi} \int_{-\infty}^{\infty} dt
	\left(
	   - \frac{1}{12}\frac{\dot D^2}{D}
	   + \frac{1}{12}\frac{1}{D}
	\right),
	\label{EffActionResult1.eq}
\end{eqnarray}
where we have redefined $aD \to D$.

\section{BACK REACTION OF THE DYNAMICAL\\ CASIMIR EFFECT}\label{back.sec}

The semiclassical effective action
for the motion of the boundaries is obtained as
\begin{eqnarray}
 S_{\rm eff}
 &=& \frac1{2\pi} \int dt
 	\left(
	       \frac{m}{2} \dot{D}^2 
	 \;-\; \frac{\kappa}{24} \frac{{\dot{D}}^2}D
	 \;+\; \frac{\kappa}{24} \frac{1}{D}
	\right),
  \label{SemiClAction.eq}
\end{eqnarray}
where $\kappa$ is the number of species of scalar fields.
The second and the third terms come from the effective action
(\ref{EffActionResult1.eq}).
In the first term we adopted the same redefinition $aD \to D$
as that in Eq. (\ref{EffActionResult1.eq}).
In this action
the second term is the back-reaction term of the dynamical Casimir
effect, and the third term is the static Casimir energy.
This action leads to the equation of motion given by
\begin{eqnarray}
 \left( m - \frac{\kappa}{12}\frac{1}{D} \right) \ddot{D}
  &=&
	- \frac{\kappa}{24} \left( \frac{\dot{D}}{D} \right)^2
	- \frac{\kappa}{24} \frac{1}{D^2}.
  \label{EquationOfMotion.eq}
\end{eqnarray}
This equation is integrable,
and the resulting relation is given by
\begin{eqnarray}
	\left( \frac m2 - \frac{\kappa}{24}\frac 1D \right) \dot{D}^2
  		        - \frac{\kappa}{24}\frac 1D
  &=&
	E,
 \label{EnergyDef.eq}
\end{eqnarray}
where $E$ is an integral constant.
The left-hand side is the Hamiltonian of this system,
thus $E$ is the energy of this system.
Here it should be noticed that
the semiclassical condition $m \gg 1/D(t)$
and the adiabatic condition $\dot{D}(t) \ll 1$
lead to the validity condition $|E| \ll m$.
Combining the equation of motion (\ref{EquationOfMotion.eq})
and the description of the energy (\ref{EnergyDef.eq}),
we obtain the mutual dynamical force between the mirrors (boundaries),
namely {\it the dynamical Casimir force},
\begin{eqnarray}
 F_{\rm dyn}
  \;\equiv\;
	m\ddot{D}
  &=&
	- \frac{\kappa}{24} \frac{1}{D^2} \;
	  \frac{1 + \dot{D}^2}
	       {1 - \frac{\kappa}{12}\frac{1}{mD}}
  \;=\;
	- \frac{\kappa}{24} \frac{1}{D^2} \;
	\frac{1 + \frac{2E}{m}}
	     {\left(1 - \frac{\kappa}{12}\frac{1}{mD}\right)^2}.
	\label{DynamicalForce.eq}
\end{eqnarray}
The dynamical Casimir force depends on
the relative velocity of the mirrors.
When the reduced mass $m$ is much larger
than the scales $E$ and $1/D$, or equivalently,
the velocity $\dot{D}$ is regarded as zero,
the dynamical Casimir force (\ref{DynamicalForce.eq}) is approximately
equal to the static one:
\begin{eqnarray}
 F_{\rm static}
  \;\equiv\;
  	- \frac{\partial}{\partial D}
	\left( - \frac{\kappa}{24} \frac{1}{D} \right)
  &=&
  	- \frac{\kappa}{24} \frac{1}{D^2}.
  \label{StaticForce.eq}
\end{eqnarray}
The ratio of the dynamical force $F_{\rm dyn}$
to the static one $F_{\rm static}$ is given by
\begin{eqnarray}
 F_{\rm dyn}/F_{\rm static}
  &=&
	\frac{1 + \dot{D}^2}
	     {1 - \frac{\kappa}{12}\frac{1}{mD}}
  \;=\;
	1 \:+\: \frac{\kappa}{12}\frac{1}{mD}
	  \:+\: \dot{D}^2
	  \:+\: \cdots.
  \label{DynamicalRatio.eq}
\end{eqnarray}
Here the $\dot{D}^2$ term in the expansion
is known as the negative-frictional-like-force~\cite{DCE_Dodonov}.
Since $\dot{D}^2 \geq 0$,
we conclude that
the dynamical force $F_{\rm dyn}$ is always attractive
and stronger than the static one $F_{\rm static}$
for $D > \frac{\kappa}{12}\frac{1}{m}$.

\section{CONCLUSION AND DISCUSSIONS}\label{summary.sec}

%
In this paper
we presented an effective theoretical approach 
to studying the Casimir effects in $1+1$-dimensions
within the adiabatic approximation.
The point of our investigation was the replacement of the spatial
configuration: $D^1 \rightarrow S^1$.
We constructed the effective action of the scalar field model,
and checked the validity of this replacement.
In our model the quantum correction to the classical kinetic term of the
mirror separation was calculated by the path-integral formalism.
The resultant quantum correction naturally
contains both the ordinary Casimir energy term and
the back-reaction term of the dynamical Casimir effect.
The semiclassical effective action (\ref{SemiClAction.eq})
was constructed of the classical kinetic term of the mirror separation
and these resultant quantum corrections.
From the action (\ref{SemiClAction.eq}),
we have obtained the dynamical vacuum pressure.
The pressure ({\it dynamical Casimir force})
includes the back-reactional force of the dynamical Casimir effect.
The dynamical Casimir force was confirmed to be attractive and always
stronger than the static Casimir force.
The dynamical Casimir force depends on
the relative velocity of the mirrors,
and it is reduced to the static one
when the velocity goes to zero.


%
The perturbative expansion of
the resultant dynamical Casimir force (\ref{DynamicalRatio.eq})
includes the term for the negative frictional force
which agrees with the result of Dodonov {\it et al.}~\cite{DCE_Dodonov}.
Although this means that our result is not entirely new,
our approach reproduces the reliable result,
thus it can be said that we have presented a unique effective
theoretical approach to the problem.


Several easier derivations of the static Casimir energy
in the Hamiltonian formulation are known,
but our method needs a more complex calculation to obtain the Casimir
energy.
%
%
Our approach, however, describes both the static and the dynamical
Casimir effects together,
and is applicable to more realistic models in 
the higher dimensions by replacing the space
$D^1 \times R^n$ into $S^1\times R^n$.

%
Furthermore, the existence of the action makes it easy for us to compare
our model with others.
For example, our model has a correspondence to
the Callan, Giddings, Harvey, Strominger (CGHS) model which describes
the two-dimensional dilaton black hole~\cite{CGHS}.
The back reaction discussed in this paper is comparable to 
the back reaction of the Hawking radiation
from the CGHS black hole~\cite{RST}.
In the CGHS model
the Hawking radiation is represented by the conformal anomaly
in the energy-momentum tensor~\cite{CGHS},
and the back reaction of the radiation, 
which is described by the Polyakov-Liouville action,
appears as the decrease in the black-hole mass~\cite{RST}.
Our classical kinetic term in the semi-classical effective action
(\ref{SemiClAction.eq})
corresponds to the kinetic term of the dilaton
in the CGHS model.

%
Some comments are in order.

%
The quantum correction (\ref{EffActionResult1.eq}) does not
include the third derivative of the dynamical variable.
This looks different from the results evaluated by 
Fulling and Davies~\cite{DCE_FD}.
They calculated the energy-momentum tensor
in (1+1)-dimensional system
of two relatively moving mirrors~\cite{DCE_FD}
as well as that in (1+1)-dimensional system
of a single non-uniformly accelerating mirror~\cite{DCE_FD,B&D}.
Both energy-momentum tensors include the third derivative
of the dynamical variables.
Our result for the system of two mirrors
does not need to coincide with
their result for the system of a single mirror
since the forms of the conformal anomaly for two systems
are different.
The result for the system of a single mirror
is due to the Unruh-like effect
rather than due to the dynamical Casimir effect.
On the other hand,
the energy-momentum tensor derived from Eq. (\ref{EffActionResult1.eq})
coincides with their result for the system of two mirrors 
under a certain transformation of the dynamical variable.

%
In the semiclassical effective action (\ref{SemiClAction.eq}),
the contribution from the dynamical Casimir effect
generated a negative-definite kinetic term of the mirror separation.
Such a kinetic term also appeared in the analysis of the CGHS
model~\cite{RST}.
The following point should be noted:
there is a positive-definite classical kinetic term,
and the negative-definite term gives only a slight correction.
This holds in the case where 
the mass scale of the mirrors $m$ is much greater
than the scale of the Casimir energy $\sim D^{-1}$.
%
%
On the other hand, 
if the mirror separation $D(t)$ is smaller
than the inverse of the mirror mass $m^{-1}$,
our result (\ref{DynamicalForce.eq}) shows that
the dynamical Casimir force $F_{\rm dyn}$ becomes repulsive.
However, our semiclassical treatment becomes unsuitable at that time.
When the motion of the mirror separation obeys the quantum mechanics,
this repulsive force might be realized.
We will leave this problem to subsequent developments.

\begin{flushleft}
 {\Large\bf ACKNOWLEDGMENTS}
\end{flushleft}

We would like to thank Professor S. Uehara, Professor M. Harada, and
Professor A. Nakayama
for their useful discussions and suggestions.
We thank Professor V. A. Miransky for discussions.
We also appreciate helpful comments of
Professor A. Sugamoto, Professor H. Funahashi,
Dr. T. Itoh, Dr. A. Takamura, Dr. S. Sugimoto,
Dr. Y. Ishimoto, and Dr. S. Yamada.
We are also grateful to Professor R. Sch\"utzhold
for his interest in our work
and enlightening discussions on related matters.
He suggested
that no particle creation occurs in our model,
and that this work is related to the dynamical back reaction of 
the static Casimir effect. 
One of us (Y.N.) is indebted to the Japan Society for
the Promotion of Science (JSPS) for its financial support.
The work is supported in part by a Grant-in-Aid for Scientific Research
from the Ministry of Education, Science and Culture (No. 03665).

\section*{APPENDIX A: Fujikawa method}
\setcounter{equation}{0}
\renewcommand{\theequation}{A\arabic{equation}}

In this appendix
we briefly explain
the derivation of the expression (\ref{EffAction2.eq})
from the definition of the effective action (\ref{EffAction1.eq}).
This derivation
is based on the evaluation of the conformal anomaly
by the Fujikawa method~\cite{Fujikawa}.
In order to perform the path integration of Eq. (\ref{EffAction1.eq}),
we make a Wick rotation
by introducing an imaginary time variable $x^2 \equiv i\eta$.
Then the Euclidean metric corresponding
to the Minkowski one (\ref{RWMetric.eq}) becomes
\begin{eqnarray}
 ds^2   &=&
	\rho(x^2) \:
	[(dx^1)^2 + (dx^2)^2].
	\label{rho.eq}
\end{eqnarray}
The Euclidean effective action is
\begin{eqnarray}
 e^{-\Gamma_E [g_{\mu\nu}]}
  &=&
  \int{\cal D}\phi\:\exp
   \left[
    - \frac1{2\pi} \int d^2x \: \sqrt{g} \: \frac12 \: g^{\mu\nu} \:
      \partial_{\mu}\phi \: \partial_{\nu}\phi
   \right].
 \label{Ea.eq}
\end{eqnarray}
By introducing $\tilde\phi\equiv\sqrt[4]{g}\:\phi$
and changing the measure ${\cal D}\phi$
into the invariant form under the general coordinate transformation
${\cal D}\tilde\phi$,
Eq. (\ref{Ea.eq}) becomes
\begin{eqnarray}
 e^{-\Gamma_E [g_{\mu\nu}]}
  &=&
   \int{\cal D}{\tilde\phi} \: \exp
   \left[ -\frac1{2\pi}
    \int d^2x \: \frac12
    \partial\left(\frac{\tilde\phi}{\sqrt\rho}\right)
    \partial\left(\frac{\tilde\phi}{\sqrt\rho}\right)
   \right].
 \label{Etea.eq}
\end{eqnarray}
Here we have used a notation
$\partial\phi\,\partial\phi
 \equiv
 \partial_1\phi\,\partial_1\phi + \partial_2\phi\,\partial_2\phi$.
We perform a mode expansion of the field $\tilde\phi(x)$
in terms of a complete set $\{\varphi_n(x)\}$:
\begin{eqnarray}
 \tilde\phi(x)
  &=&        \sum_n \: a_n \varphi_n(x) 
  \;\equiv\; \sum_n \: \langle x | n \rangle \: a_n,
\end{eqnarray}
where we have chosen ${\varphi_n(x)}$
as an eigenfunction of the Hamiltonian operator,
\begin{eqnarray}
 \hat{H}
  &=& -\frac12\,\frac1{\sqrt{\rho}}\,\partial\partial\,\frac1{\sqrt{\rho}},
  \qquad
  \hat{H} \varphi_n(x)
  \;=\; \lambda_n^2 \, \varphi_n(x).
\label{Ham.eq}
\end{eqnarray}
Here $\varphi_n(x)$ satisfies the normalization
$\int d^2x\:\varphi^{\daggerC}_m(x)\:\varphi_n(x) \:=\: \delta_{mn}$.
Now
we note that the measure ${\cal D}\tilde\phi$ is expressed
by the mode coefficients $a_n$ as
\begin{eqnarray}
 {\cal D}\tilde\phi
  &=&   \prod_x \: {\cal D}\tilde\phi(x)
  \;=\; \left[\,\det\langle x|n\rangle\,\right] \:
        \prod_n \: da_n
  \;=\; \prod_n \: da_n.
\end{eqnarray}

Under the Weyl transformation
$g_{\mu\nu}\to e^{2\alpha(x)}g_{\mu\nu}$
the mode coefficients of the field $\tilde\phi(x)$, $a_n$,
are transformed as an infinitesimal form,
\begin{eqnarray}
 \tilde\phi(x)
  &\to&
	\tilde\phi^{\prime}(x)
  \;\equiv\;
	\sum_n \, a^{\prime}_n \, \varphi_n(x),
  \nonumber\\
 a^{\prime}_n
  &=&
 a_n \:+\: \sum_m \frac12 \int d^2x \:
	\alpha(x) \varphi^{\daggerC}_n(x) \: \varphi_m(x)a_m
  \;\equiv\;
	\sum_m C_{nm} \, a_m.
\end{eqnarray}
Then the measure is transformed as
\begin{eqnarray}
 {\cal D}\tilde\phi^{\prime}
  &=&
	\prod_n da^{\prime}_n
  \;=\;
	\left[\det (C_{nm})\right] \: \prod_l da_l
  	\nonumber\\
  &=&
	\exp \left[ \Tr\ln
	      \left(
	       \delta_{nm}
	       \;+\; \frac12\int d^2x\:\alpha(x)
                     \varphi^{\daggerC}_n(x) \: \varphi_m(x)
	      \right)
	     \right]
	\; \prod_l da_l
	\nonumber\\
 &=&
	\exp \left[+\frac12\int d^2x\:\alpha(x)
	      \sum_n\varphi^{\daggerC}_n(x)\:\varphi_n(x)
	     \right] \; {\cal D}\tilde{\phi}.
\end{eqnarray}
This gives the Jacobian of the conformal transformation.
By the Weyl transformation chosen $\alpha(x)=-\frac12\ln\rho(x)$
for $\tilde{\phi} \to \tilde{\phi}^{\prime} = \tilde{\phi} / \sqrt{\rho}$,
the effective action (\ref{Etea.eq}) becomes
\begin{eqnarray}
 e^{-\Gamma_E [g_{\mu\nu}]}
 &=& \exp\left[- \frac12 \int d^2x \, \alpha(x)
               \sum_n \varphi^{\daggerC}_n(x) \, \varphi_n(x)
         \right] \nonumber\\
 & &  \times
     \int{\cal D}{\tilde\phi^\prime} \,
     \exp\left[- \frac1{2\pi}
	  	 \int d^2x \, \frac12 \,
		 \partial\tilde\phi^\prime \,
                 \partial\tilde\phi^\prime
         \right],
\label{EffActionE3.eq}
\end{eqnarray}
where the second factor equals to the
partition function of the free scalar field in the flat space time.
Finally, we can arrive at our destination (\ref{EffAction2.eq})
from the description (\ref{EffActionE3.eq})
by the inverse Wick rotation.

\section*{APPENDIX B: Another path-integral Calculation of the Casimir energy}
\setcounter{equation}{0}
\renewcommand{\theequation}{B\arabic{equation}}

In this appendix we give another partition-functional derivation of
the Casimir energy by means of the
point-splitting ansatz and the Feynman prescription:
In the path-integral method
the partition function part in (\ref{EffAction2.eq})
can be also evaluated
by using the point-splitting ansatz 
and the Feynman's renormalization prescription.
Employing the ansatz of point splitting to (\ref{partition.eq}),
\begin{eqnarray}
 Z_E &=& \int{\cal D}{\phi}\,
  \exp \left[
	-\frac{1}{2\pi}\int d^2x \frac{1}{2\pi}\int d^2x^{\prime}
	\frac{1}{2} \phi(x)A(x,x^{\prime})\phi(x^{\prime})
	\right] \nonumber\\
   &=& \exp \left[-\frac1{2}\Tr\ln A \right],
\end{eqnarray}
where $\int d^2x \equiv \int_{-\infty}^{\infty} dx^2 \int_0^a dx^1$, and
$ A(x,x^{\prime}) \equiv
\delta^{\mu\nu}\partial_{\mu}\partial^{\prime}_{\nu}
\delta^{(2)}(x-x^{\prime})$.
The two-dimensional Dirac delta function in the integral
representation is
\begin{eqnarray}
 \delta^{(2)}(x-x^{\prime}) &=& 
  \frac1{a} \,\sum_{n=-\infty}^{+\infty}
  \int \frac{dk}{2\pi} \, e^{-ik(x^2-x^{\prime 2})}
  \, e^{i\frac{2\pi n}{a}(x^1-x^{\prime 1})} \; .
\end{eqnarray}
Now we come back to Minkowski space 
and introduce mass $M$ of the scalar field
to regularize the integral,
\begin{eqnarray}
 \frac{1}{i} \ln Z 
  &=& 
  - \frac{1}{2} \,
  \frac{1}{(2\pi)^2}\int d^2x \int d^2x^{\prime} \,
  \delta^{(2)}(x-x^{\prime}) \,
  \frac1a \, \sum_n \,
  e^{i\frac{2\pi n}{a}(x-x^{\prime})} \nonumber\\
  && \times \; \int dM^2
  \int \frac{dk}{2\pi} \,
  \frac{e^{-ik(\eta-\eta^{\prime})}}
  {-k^2 + \left(2\pi n / a\right)^2 + M^2} \;.
\end{eqnarray}
With the $i\epsilon$ prescription,
we perform the integral in the complex $k$ plane,
applying the residue theorem,
\begin{eqnarray}
 && \int \frac{dk}{2\pi} \,
  \frac{e^{-ik(\eta-\eta^{\prime})}}
  {-k^2 + \left(2\pi n / a\right)^2 + M^2  - i\epsilon} \nonumber\\
 && = \theta(\eta - \eta^{\prime})
  \frac{i}{2\sqrt{\left(2\pi n / a\right)^2 + M^2}}
  + \theta(\eta^{\prime} - \eta)
  \frac{-i}{-2\sqrt{\left(2\pi n / a\right)^2 + M^2}},
\end{eqnarray}
where $\epsilon>0$ and $\theta(x)$ is the step function.
Here we define the new parameter,
$\omega_n^2 \equiv \left(2\pi n / a\right)^2 + M^2$,
and replace the integral into the following form:
\begin{eqnarray}
 \int dM^2 \, \frac 1{2\sqrt{\left(2\pi n / a\right)^2 + M^2}}
  &=& \int d\omega_n
 \;=\; \omega_n.
\end{eqnarray}
Then we take the massless limit $M\to 0$ and perform the summation,
\begin{eqnarray}
 \sum_{n=-\infty}^{+\infty} \omega_n
   &=& \frac{4\pi}{a}\sum_{n=1}^{\infty} n
  \;=\; \frac{4\pi}{a}\zeta(-1)
  \;=\; -\frac{\pi}{3a}.
\end{eqnarray}
At last we arrive at the same form of Eq. (\ref{PartitionFuncResult.eq}),
\begin{eqnarray}
 \frac1{i}\ln Z
  \;=\; \frac 1{2\pi} \int_{-\infty}^{\infty}dt \;
  \frac 1{12} \frac 1{aD(t)}.
\end{eqnarray}

\newcommand{\PRL}[3]	{{Phys. Rev. Lett.}   {\bf #1}, #2 (#3)}
\newcommand{\PR}[3]	{{Phys. Rev.}         {\bf #1}, #2 (#3)}
\newcommand{\PRA}[3]	{{Phys. Rev. A}       {\bf #1}, #2 (#3)}
\newcommand{\PRD}[3]	{{Phys. Rev. D}       {\bf #1}, #2 (#3)}
\newcommand{\PL}[3]	{{Phys. Lett.}        {\bf #1}, #2 (#3)}
\newcommand{\PLA}[3]	{{Phys. Lett. A}      {\bf #1}, #2 (#3)}
\newcommand{\PLB}[3]	{{Phys. Lett. B}      {\bf #1}, #2 (#3)}
\newcommand{\NuP}[3]	{{Nucl. Phys.}        {\bf #1}, #2 (#3)}
\newcommand{\PTP}[3]	{{Prog. Theor. Phys.} {\bf #1}, #2 (#3)}
\newcommand{\Nature}[3]	{{Nature (London)}    {\bf #1}, #2 (#3)}
\newcommand{\PKNAW}[3]	{{Proc. K. Ned. Akad. Wet.}     {\bf #1}, #2 (#3)}
\newcommand{\Physica}[3]{{Physica (Utrecht)}  {\bf #1}, #2 (#3)}
\newcommand{\JMP}[3]	{{J. Math. Phys.}     {\bf #1}, #2 (#3)}
\newcommand{\PRSLA}[3]	{{Proc. R. Soc. London, Ser A}   {\bf #1}, #2 (#3)}
\newcommand{\AP}[3]	{{Ann. Phys. (N.Y.)}  {\bf #1}, #2 (#3)}
\newcommand{\JPA}[3]	{{J. Phys. A}          {\bf #1}, #2 (#3)}
\newcommand{\ZhETF}[3]	{{Zh. \'{E}ksp. Teor. Fiz. Pis'ma. Red.}  {\bf #1}, #2 (#3)}
\newcommand{\JETP}[3]	{{JETP Lett.}         {\bf #1}, #2 (#3)}
\newcommand{\CMP}[3]	{{Commun. Math. Phys.} {\bf #1}, #2 (#3)}


\end{document}